\documentclass[aps,pra,prabib,epsfig,amsmath,amssymb,twocolumn]{revtex4}
\usepackage{bbold}
\usepackage{verbatim}
\usepackage{amssymb}
\usepackage{amsmath}
\usepackage{graphicx}
\usepackage{subfigure}
\usepackage{dcolumn}
\usepackage{bm}

\newcommand{\sgn}{\operatorname{sgn}}

\begin{document}


\title{Short range asymptotic behavior of the wave-functions of interacting spin-half fermionic atoms with spin-orbit coupling: a model study}


\author{Yuxiao Wu}
\author{Zhenhua Yu}
\email{huazhenyu2000@gmail.com}
\affiliation{Institute for Advanced Study, Tsinghua University, Beijing 100084, China
}%


\date{\today}


\begin{abstract}
We consider spin-half fermionic atoms with isotropic Rashba spin-orbit coupling in three directions. The interatomic potential is modeled by a square well potential. We derive the analytic form of the asymptotic wave-functions at short range of two fermions in the subspace of zero net momentum and zero total angular momentum. We show that the spin-orbit coupling has perturbative effects on the short range asymptotic behavior of the wave-functions away from resonances. We argue that our conclusion should hold generally.
\end{abstract}

\maketitle


\section{Introduction}
The study of dilute unitary atomic Fermi gases has significantly and substantially extended our knowledge on
strongly interacting many-body systems \cite{stringari}. Theoretic advances in this direction have been propelled by observing
the short-range asymptotic behavior of the wave-function of the systems \cite{tan,braaten, shizhong}. 
In such ultra-cold dilute gases, when the separation $r$ between two fermions of different species is much smaller than the mean inter-particle spacing $d$, but bigger than $r_0$, the range of the interatomic potential, the pair wave-function of the two fermions is mainly $s$-wave and has the asymptotic form $\psi_s(r)=1-a_s/r$; when a similar situation happens to two fermions of same species, the pair wave-function is mainly $p$-wave and its radial part has the asymptotic form $\psi_p(r)=k(r-u_p/r^2)$. Here $k$ is the relative wave vector between the two fermions and of order $1/d$, $a_s$ is the $s$-wave scattering length and $u_p$ is the $p$-wave scattering volume. Thus, when considering the short-range correlations, the $p$-wave part can be safely neglected compared to the $s$-wave part unless $u_p$ becomes divergent. Based on this observation, a collection of remarkable relations regarding the short-range correlations and various physical observables have been derived \cite{tan,braaten, shizhong}.

On the other hand, recent successful engineering of synthetic gauge field adds another important dimension to atomic gases \cite{ianprl, iannat, iannp,iannat2, zhang, martin}. Raman processes couple different hyper-fine states of atoms and realizes a model with an effective spin-orbit coupling for spin-half particles when the hyper-fine states other than the lowest two are adiabatically eliminated. This success raises the prospect of using ultra-cold atomic systems to study and simulate spin-orbit coupling physics \cite{dalibard}, which has excited great interest in condensed matter physics \cite{kane,qi}. In the context of Fermi gases, subsequent theoretical studies investigated the effects of the spin-orbit coupling on the scattering and bound states of two fermionic atoms \cite{vijay2, cui, galitski,vijay,peng,peng3,peng2}, the BEC-BCS crossover in two and three dimensions \cite{vijay0,chuanwei3,hu,zhai,jiang,carlos2,chuanwei2,huang2,huang3}, collective motions in the fermionic superfluid phase \cite{vijay3,iskin2,iskin3,huang}, the equation of state in the high temperature regime \cite{zhenhua}, and its joint effects with polarization \cite{guo,iskin,yi,liu,wan,carlos3} and mass imbalance \cite{iskin}, and possible emergence of majorana fermions \cite{chuanwei,carlos}.

The introduction of the spin-orbit coupling to ultracold atomic gases also raises another interesting question: How would the short-range asymptotic behavior of the wave-function of atomic gases be modified? Attempts to answer this question include: Cui studied the two-body problem of fermions with symmetric Rashba spin-orbit coupling in three directions and discovered that the usual $s$-wave pseudo-potential is still applicable in certain regimes. This indicates that the effects of the spin-orbit coupling on the short-range asymptotic behavior of the wave-function can be perturbative \cite{cui}. Later Ref.~\cite{zhenhua} presented a general argument based on the magnitudes of relevant length scales. Since the spin-orbit coupling strength, as usually realized in the experiments by Raman processes, corresponds to a length scale $\sim500$ nm, which is much larger than, e.g., the interatomic potential range $\sim50$ nm, the wave-function inside the range of the interatomic potential should remain intact to zero order. The asymptotic behavior of the wave-function outside the interatomic potential range which is determined by the one inside the range should stay unchanged as well. Reference \cite{peng} reached a similar conclusion by an argument in which unitary transformations are implemented to eliminate the spin-orbit coupling from the kinetic part of the Hamiltonian. 
However, besides the case with spin-orbit coupling in one direction \cite{peng}, the general arguments mentioned above await justification from explicitly worked out examples. 

In this paper, we study the problem of two spin-half fermions with isotropic spin-orbit coupling in three directions. The single-particle Hamiltonian of the fermions is
\begin{equation}
H_1=\frac{\mathbf p^2}{2m}+\frac{\lambda}{m}\mathbf p\cdot\boldsymbol\sigma+\frac{\lambda^2}{2m}
\end{equation}
with $m$ the mass of the atoms. Without loss of generality, we assume the spin-orbit coupling strength to be positive, $\lambda>0$. 
Possible experimental realization of the spin-orbit coupling of the form $\mathbf p\cdot\boldsymbol\sigma$ has been proposed in Ref.~\cite{spielman}.
We model the interatomic potential by an attractive square well potential, $V(r)=-V_0\theta(r_0-r)$, where $r_0$ is the potential range and $V_0>0$. As of experimental interest, we assume $\lambda r_0\ll1$. 
We calculate the scattering and bound state wave-functions of the two interacting fermions in the subspace of zero net momentum and zero total angular momentum (sum of the spin and orbital angular momentum). We derive analytically the asymptotic behavior of the wave-functions outside the range of the square well potential. Away from resonances, the modification of the asymptotic behavior due to the spin-orbit coupling is shown to be perturbative. We explain this perturbative effect by inspecting the wave-functions inside the potential range and argue that our conclusion is valid for generic situations with spin-orbit coupling.

\section{the Schr\"odinger equation}
The Hamiltonian of two interacting fermions can be cast into the form
\begin{align}
H_2=&H_K+H_k\\
H_K=&\frac{\mathbf K^2}{4m}+\frac{\lambda\mathbf K}{2m}\cdot(\boldsymbol\sigma_2+\boldsymbol\sigma_1)\label{hkk}\\
H_k=&\frac{\mathbf k^2}{m}+\frac{\lambda\mathbf k}{m}\cdot(\boldsymbol\sigma_2-\boldsymbol\sigma_1)+\frac{\lambda^2}{m}+V(r).\label{hk}
\end{align}
Here $\mathbf K$ is the net momentum of the two fermions and $\mathbf k$ is the relative one, and $\mathbf r$ are the relative coordinates. The subscript of $\boldsymbol\sigma$ labels for the $i$th-fermion. From Eqs.~(\ref{hkk}) and (\ref{hk}), the motion of the center of mass and the relative one are coupled together via the spin operators. In the following, we focus on the subspace of $\mathbf K=0$ and $\mathbf J\equiv\mathbf L+\mathbf S=0$, with $\mathbf L$ the orbital angular momentum and $\mathbf S$ the total spin, where analytic results can be derived.

The relative wave-function $\Psi(\mathbf r)$ in the subspace $\mathbf K=0$ satisfies the Schr\"odinger equation
\begin{equation}
\left[\frac{\hat{\mathbf{k}}^2+\lambda^2}{m}+M_\lambda+V(r)\right]\Psi(\mathbf r)=E\Psi(\mathbf r),\label{sch}
\end{equation}
with
\begin{align}
M_\lambda&=2\frac{\lambda}{m}\begin{bmatrix} 0 & \frac{\hat{k}_x+i\hat{k}_y}{\sqrt{2}} & -\frac{\hat{k}_x-i\hat{k}_y}{\sqrt{2}} & -\hat{k}_z\\
\frac{\hat{k}_x-i\hat{k}_y}{\sqrt{2}} & 0 &0 & 0\\
-\frac{\hat{k}_x+i\hat{k}_y}{\sqrt{2}} & 0 & 0& 0\\
-\hat{k}_z & 0 & 0 & 0\\\end{bmatrix}
\end{align}
when represented in the spin basis $[(\uparrow\downarrow-\downarrow\uparrow)/\sqrt2, \uparrow\uparrow, \downarrow\downarrow, (\uparrow\downarrow+\downarrow\uparrow)/\sqrt2)]$ for the two spin-half fermions. Since each elements of $M_\lambda$ are proportional to the spherical harmonic functions $Y_{1,m}(\Omega_{\mathbf k})$ or their complex conjugates, with further contraint $\mathbf J=0$ we have 
\begin{equation}
\Psi(\mathbf r)=\psi_0(r)\begin{bmatrix}Y_{0,0}(\Omega_{\mathbf r})\\0\\0\\0\end{bmatrix}
-\frac i{\sqrt3}\psi_1(r)\begin{bmatrix}
0\\Y_{1,-1}(\Omega_{\mathbf r})\\ 
Y_{1,1}(\Omega_{\mathbf r})\\ - Y_{1,0}(\Omega_{\mathbf r})\end{bmatrix}.
\end{equation}
The new spinor wave-function $\Phi(r)=[\psi_0(r),\psi_1(r)]^T$ satisfies
\begin{align}
&\left\{-\frac{1}{r^2}\frac{d}{d r}\left(r^2\frac{d}{d r}\right)+\lambda^2+m[V(r)-E]\right.\nonumber\\
&\left.+\begin{bmatrix} 0 & -2\lambda\left(d/dr+2/r\right)\\
2\lambda \left(d/dr\right) & 2/r^2 \end{bmatrix}\right\}\Phi(r)=0.\label{sch2}
\end{align}

\section{scattering states}
It is straightforward to show that for the scattering states, the wavefunctions are
\begin{align}
\Phi^<(r)=
A\begin{bmatrix}j_0(q_1r)\\j_1(q_1r) \end{bmatrix}+B\begin{bmatrix}j_0(q_2r)\\j_1(q_2r)\end{bmatrix}\label{in}
\end{align}
for $r<r_0$, and
\begin{align}
\Phi^>(r)=
&C \begin{bmatrix}h^{(2)}_0(k_1r)\\ -h^{(2)}_1(k_1r) \end{bmatrix}+D \begin{bmatrix}h^{(1)}_0(k_1r)\\-h^{(1)}_1(k_1r) \end{bmatrix}\nonumber\\
+&E\begin{bmatrix}h^{(1)}_0(k_2r)\\h^{(1)}_1(k_2r) \end{bmatrix}+F\begin{bmatrix}h^{(2)}_0(k_2r)\\h^{(2)}_1(k_2r) \end{bmatrix}
\end{align}
for $r>r_0$.
Here $j_i$ are the spherical Bessel functions and $h^{(1,2)}_i$ are the spherical Hankel functions, and
\begin{align}
 k_{1}\equiv k-\lambda,\,\, &k_2\equiv k+\lambda,\\
 q_1\equiv\lambda-q,\,\, &q_2\equiv\lambda+q, \\
 k\equiv \sqrt{mE},\,\, & q\equiv\sqrt{m(E+V_0)}.
\end{align}
Since there are six coefficients ($A$ to $F$) and four connection conditions at $r=r_0$ for the wavefunctions and an overall normalization factor, the wavefunctions can be determined up to a coefficient; there are two linearly independent solutions for each energy $E$. Let us choose the $i$th linearly independent solution $\Phi_i$ as $\Phi_i^<=[j_0(q_ir),j_1(q_ir)]^T$ for $r<r_0$. By requiring $\Phi_i$ and its first derivative continuous at $r=r_0$, we find the corresponding coefficients for $\Phi_i^>$
\begin{align}
C_i=&\frac{k_2-q_i}{k_1+k_2}
\frac{h^{(1)}_0(\tilde k_1)j_1(\tilde q_i)+h^{(1)}_1(\tilde k_1)j_0(\tilde q_i)}{h^{(1)}_1(\tilde k_1)h^{(2)}_0(\tilde k_1)-h^{(1)}_0(\tilde k_1)h^{(2)}_1(\tilde k_1)}\label{c}\\
E_i=&\frac{k_1+q_i}{k_1+k_2}\frac{h^{(2)}_0(\tilde k_2)j_1(\tilde q_i)-h^{(2)}_1(\tilde k_2)j_0(\tilde q_i)}{h^{(2)}_0(\tilde k_2)h^{(1)}_1(\tilde k_2)-h^{(2)}_1(\tilde k_2)h^{(1)}_0(\tilde k_2)}\label{e}\\
D_i=&C_i^*\\
F_i=&E_i^*,
\end{align}
with $\tilde k_i\equiv k_i r_0$ and $\tilde q_i\equiv q_i r_0$.

The physical meaning of the above coefficients can be understood in the following way. Without interaction, $V(r)=0$, either $[j_0(k_1r),j_1(k_1r)]^T$ or $[j_0(k_2r),j_1(k_2r)]^T$ is the free particle solution to Eq.~(\ref{sch2}). The interaction $V(r)$ realizes mutual scattering between the two waves. We can construct two new linearly independent solutions as
\begin{align}
\Phi_{k_1}^>
=&\frac{E^*_2\Phi_1^>-E^*_1\Phi_2^>}{E^*_2C_1-E_1^*C_2}
=\begin{bmatrix}h^{(2)}_0(k_1 r)\\ -h^{(2)}_1(k_1 r)\end{bmatrix}\nonumber\\&+S_{11}\begin{bmatrix}h^{(1)}_0(k_1 r)\\ - h^{(1)}_1(k_1 r)\end{bmatrix}+S_{21}\frac{k_2}{k_1}\begin{bmatrix}h^{(1)}_0(k_2 r)\\ h^{(1)}_1(k_2 r)\end{bmatrix}\\
\Phi_{k_2}^>
=&\frac{C_2\Phi_1^>-C_1\Phi_2^>}{C_2E_1^*-C_1E_2^*}
=\begin{bmatrix}h^{(2)}_0(k_2 r)\\ h^{(2)}_1(k_2 r)\end{bmatrix}\nonumber\\&+S_{22}\begin{bmatrix}h^{(1)}_0(k_2 r)\\ h^{(1)}_1(k_2 r)\end{bmatrix}
+S_{12}\frac{k_1}{k_2}\begin{bmatrix}h^{(1)}_0(k_1 r)\\ -h^{(1)}_1(k_1 r)\end{bmatrix},
\end{align}
where $S_{ij}$ are the elements of the matrix
\begin{align}
S=&\frac1{E_2^*C_1-E_1^*C_2}\nonumber\\
\times&\begin{bmatrix} E_2^*C_1^*-E_1^*C_2^* & \frac{k_2}{ k_1}(C_2^*C_1-C_1^*C_2)\\
\frac{k_1}{ k_2}(E_2^* E_1-E_1^*E_2) & C_1E_2-C_2E_1\end{bmatrix}.\label{smatrix}
\end{align}
Since $h^{(1)}_i(x)\sim e^{ix}/x$ and $h^{(2)}_i(x)\sim e^{-ix}/x$ when $x\to\infty$, the wave-function $\Phi^>_{k_1}$($\Phi^>_{k_2}$) describes the process that the incoming wave of wave vector $k_1$($k_2$) is scattered by $V(r)$ into the outgoing waves of wave vectors $k_1$ and $k_2$. The matrix $S$ formed by the coefficients $S_{ij}$ is the scattering $S$-matrix. It is straightforward to show that $S$ satisfies the unitary condition  $S^\dagger S=1$. The appearance of the ratios between $k_1$ and $k_2$ in Eq.~(\ref{smatrix}) is because the two wave vectors are of different magnitude. 

By diagonalizing the unitary $S$-matrix, we obtain the standing-wave solutions
\begin{align}
&\Phi_{\pm}^>\nonumber\\
=&v^\pm_1k_1\Phi^>_{k_1}+v^\pm_2k_2\Phi^>_{k_2}\nonumber\\
=&v^\pm_1k_1 \begin{bmatrix}h^{(2)}_0(k_1 r)\\ -h^{(2)}_1(k_1 r)\end{bmatrix}+v^\pm_2k_2\begin{bmatrix}h^{(2)}_0(k_2 r)\\ h^{(2)}_1(k_2 r)\end{bmatrix}\nonumber\\
&+e^{2i\delta_\pm}\left\{v^\pm_1k_1 \begin{bmatrix}h^{(1)}_0(k_1 r)\\ -h^{(1)}_1(k_1r)\end{bmatrix}+v^\pm_2k_2\begin{bmatrix}h^{(1)}_0(k_2 r)\\ h^{(1)}_1(k_2 r)\end{bmatrix}\right\},\label{phiv}
\end{align}
where $[v^\pm_1,v^\pm_2]^T$ are the eigenvectors of $S$ given by Eq.~(\ref{smatrix}) and $e^{2i\delta_\pm}$ are the corresponding eigenvalues. From Eqs.~(\ref{c}) and (\ref{e}), one can prove $k_1^2(E_2^*E_1-E_1^*E_2)=k_2^2(C_2^*C_1-C_1^*C_2)$. Since the $S$-matrix can be expanded as $S=\tilde S_0\mathbb1+\tilde S_x\sigma_x+\tilde S_z\sigma_z$ and the ratio between the coefficients $\tilde S_x$ and $\tilde S_z$ is real, we can choose the eigenvectors $[v^\pm_1,v^\pm_2]^T$ to be real. Note that $[\Phi_{\pm}^>]^*$ is proportional to $\Phi_{\pm}^>$ apart from a phase. This complies with the expectation that solutions to the one-dimensional differential equation (\ref{sch2}) can be chosen to be real.
Equation (\ref{phiv}) corresponds to the ansatz used in Ref.~\cite{cui} [cf.~Eqs.~(31) and (32) therein]. 
The two new phase shifts $\delta_\pm$, characterizing the scattering effects, are the counterparts of the $s$-wave and $p$-wave phase shifts $\delta_s$ and $\delta_p$ in the absence of the spin-orbit coupling.

\begin{figure}
\includegraphics[width=7cm]{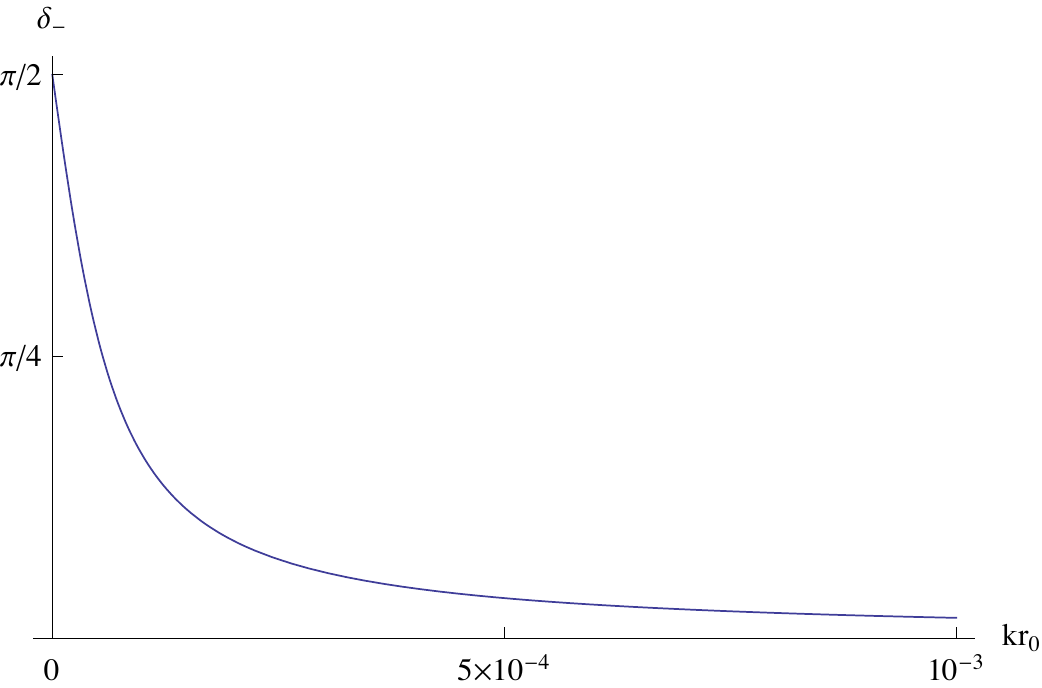}
\caption{Phase shift $\delta_-(kr_0)$ vs $kr_0$ for $\tilde\lambda=0.01$ and $\eta=1$.} \label{minus}
\end{figure}

\begin{figure}
\includegraphics[width=7cm]{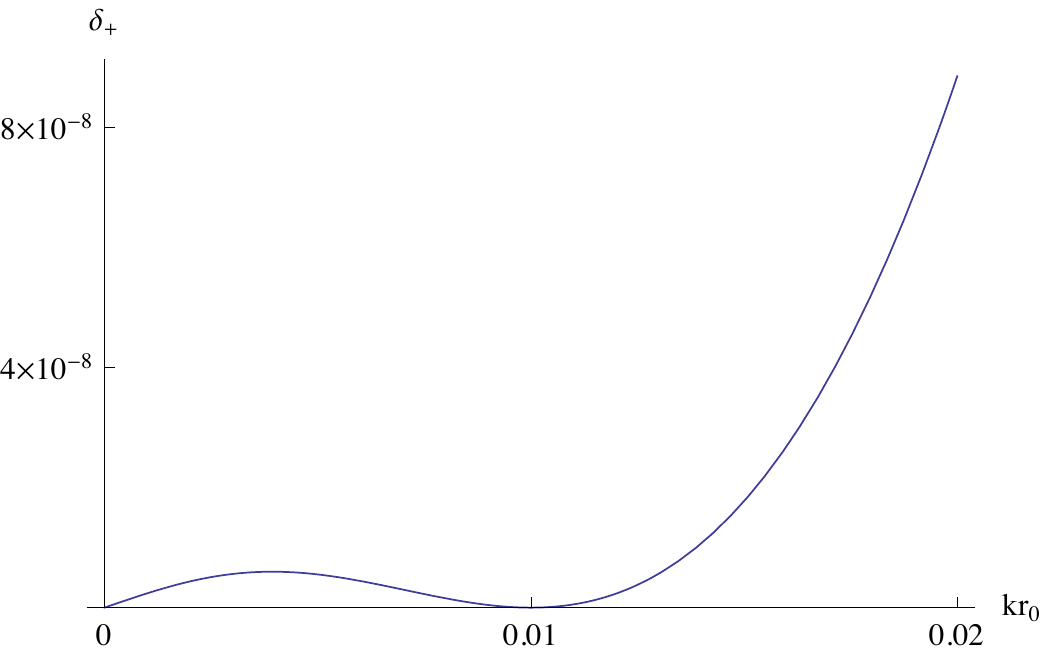}
\caption{Phase shift $\delta_+(kr_0)$ vs $kr_0$ for $\tilde\lambda=0.01$ and $\eta=1$.} \label{plus}
\end{figure}

We plot $\delta_-(kr_0)$ in Fig.~(\ref{minus}) and $\delta_+(kr_0)$ in Fig.~(\ref{plus}) for $\tilde\lambda=0.01$ and $\eta=1$ with $\tilde\lambda\equiv\lambda r_0$ and $\eta\equiv\sqrt{mV_0r_0^2}$. Generically we find $\exp[2i\delta_-(0)]=-1$ and $\exp[2i\delta_+(0)]=1$. The unitarity of $\delta_-(0)$ is a reminiscent of one-dimensional scattering \cite{cui}. When both $\delta_-$ and $\delta_+$ are defined in the domain $[-\pi/2,\pi/2]$, to zero order of $\tilde\lambda$, $\delta_-(0)$ jumps by $\pi$ where $a_s$ switches between $-\infty$ to $+\infty$, and the slope of $\delta_-(k)$ in the small $k$ limit changes sign where $u_p$ flips between $-\infty$ to $+\infty$.
To analyze the properties of $\delta_\pm$, we expand the $S$-matrix to first order of $k$ 
\begin{align}
S=\sigma_x+M\tilde k+\mathcal O(\tilde k^2),
\end{align}
where $\tilde k\equiv kr_0$. The elements $M_{ij}$ of the matrix $M$ are listed in the Appendix.
Correspondingly in such limit, we have $\delta_-=-\sgn(\delta_-')\pi/2+\delta_-'\tilde k$, $\delta_+=\delta_+'\tilde k$ and
\begin{align}
\mathbf v^-=&\frac1{\sqrt2}\left(\begin{bmatrix}-1\\1\end{bmatrix}+\delta v\tilde k \begin{bmatrix}1\\1\end{bmatrix}\right),\\
\mathbf v^+=&\frac1{\sqrt2}\left(\begin{bmatrix}1\\1\end{bmatrix}-\delta v \tilde k\begin{bmatrix}-1\\1\end{bmatrix}\right),
\end{align}
with
\begin{widetext}
\begin{align}
\delta_-'=&\frac i4(M_{11}+M_{22}-M_{12}-M_{21})\nonumber\\
=&\{2 \tilde\lambda^4 -2 \eta^2 \tilde\lambda^2 +
    \eta [\eta^2  + \tilde\lambda^2 - 2 \eta^2  \tilde\lambda^2 + 
      2 \tilde\lambda^4 + (\eta^2  + \tilde\lambda^2) \cos(2 \tilde\lambda)] \sin(
     2  \eta)+ 
   2 \eta^2  \tilde\lambda [\tilde\lambda \cos(2\tilde\lambda) - \sin(2 \tilde\lambda)]  \nonumber\\
&  + 2 \tilde\lambda \cos(
     2  \eta) [-2\eta^2  \tilde\lambda + \tilde\lambda^3 - 
     \eta^2  \sin(2 \tilde\lambda)]\}/\{2 \eta^2  \tilde\lambda^2 [\cos(2  \eta) - 
      \cos(2 \tilde\lambda)] + 
   2  \eta \tilde\lambda^2 (\eta^2  - \tilde\lambda^2) \sin(2  \eta)\},
\end{align}
\begin{align}
\delta_+'=&-\frac {i}4(M_{11}+M_{22}+M_{12}+M_{21})\nonumber\\
=&\{\eta [\eta^2 + \tilde\lambda^2 - 2\eta^2 \tilde\lambda^2 + 
      2\tilde\lambda^4 - (\eta^2 + \tilde\lambda^2) \cos(2 \tilde\lambda)] \sin(
     2 \eta)+ 
   2 \tilde\lambda \cos(
     2 \eta) [ \tilde\lambda^3 -2 \eta^2 \tilde\lambda +\eta^2 \sin(2 \tilde\lambda)]\nonumber\\
     & - 2 \tilde\lambda[ \tilde\lambda^3-\eta^2 \tilde\lambda - 
      \eta^2 \tilde\lambda \cos(2 \tilde\lambda) + 
      \eta^2 \sin(2\tilde\lambda)]\}/
     \{2 \eta^2  \tilde\lambda^2 [\cos(2  \eta) - 
      \cos(2 \tilde\lambda)] + 
   2  \eta \tilde\lambda^2 (\eta^2  - \tilde\lambda^2) \sin(2  \eta)\},
   \end{align}
   \begin{align}
\delta v=&(M_{11}-M_{22})/4\nonumber\\
=&\{2 \eta \tilde\lambda[\cos(2 \eta) \cos(2 \tilde\lambda)-1 ] + 
  (\eta^2 + \tilde\lambda^2) \sin(2 \eta) \sin(2\tilde\lambda)\}/\{
 2 \eta\tilde\lambda^2  [\cos(2 \eta)- \cos(2\tilde\lambda)] + 2\tilde\lambda^2(\eta^2 -\tilde\lambda^2) \sin(
      2 \eta)\}.
\end{align}
\end{widetext}
The three quantities, $\delta_-'$, $\delta_+'$ and $\delta v$, diverge when 
\begin{align}
\frac{\eta}{\eta-\tilde\lambda}[\cos(2\eta)-\cos(2\tilde\lambda)]+(\eta+\tilde\lambda)\sin(2\eta)=0,\label{bcon}
\end{align}
where a bound state emerges [cf.~Eq.~(\ref{long})]. This divergence reflects that bound state energies are the poles of the $S$-matrix. Away from the singularities, these quantities have the asymptotic expansions in $\tilde\lambda$ as
\begin{align}
\delta_-'
=&\frac{r_0}{\tilde\lambda^2 a_s}\left[1+\mathcal O(\tilde\lambda^2)\right],\label{dminusa}\\
\delta_+'
=&-\frac{\lambda^2 u_p}{3r_0}\left[1+\mathcal O(\tilde\lambda^2)\right],\label{dplusa}\\
\delta v
=&\left[\frac{1}{\tilde \lambda}-\frac{2\lambda u_p}{3 a_s r_0}\right]+\mathcal O(\tilde\lambda^3)\label{dva}.
\end{align}
For our square well potential model, the $s$-wave scattering length $a_s$ is given by $r_0/a_s=\eta/(\eta-\tan\eta)$, and the $p$-wave scattering volume $u_p$ by $u_p/r_0^3=1-3(1-\eta\cot\eta)/\eta^2$. Equations (\ref{dminusa}) and (\ref{dplusa}) explain the jumps of $\delta_-(0)$ and the sign change of the slope of $\delta_+(k)$ for small $k$.

Exceptions occur at resonances: at the first set of resonances given by Eq.~(\ref{cri}), we find $\delta_-(0)=-\delta_+(0)\approx-\tilde\lambda/2$, $\mathbf v^-=[0,1]^T$, and $\mathbf v^+=[1,0]^T$; at the second set, $\delta_-(0)=-\delta_+(0)\approx2\tilde\lambda^3/3$, $\mathbf v^-=[0,1]^T$, and $\mathbf v^+=[1,0]^T$.

The asymptotic form of Eq.~(\ref{phiv}) in the region $r\gtrsim r_0$ is
\begin{align}
\Phi^>_{\pm}=&
\left\{\frac{v^\pm_1+v^\pm_2}r+(v^\pm_1k_1+v^\pm_2k_2)\cot\delta_\pm \right\}\begin{bmatrix}1\\0\end{bmatrix}\nonumber\\
&+\left\{\left(\frac{v^\pm_2}{k_2}-\frac{v^\pm_1}{k_1}\right)\frac1{r^2}+\frac12(v^\pm_2 k_2-v^\pm_1 k_1)\right.\nonumber\\
&\left.+\frac13\cot\delta_\pm(k_2^2 v^\pm_2-k_1^2 v^\pm_1) r \right\}\begin{bmatrix}0
\\1\end{bmatrix}\label{asymp}.
\end{align}
Note that an overall factor $2e^{\delta_\pm}\sin\delta_\pm$ has been omitted. 
In the low energy limit $k\to0$, when away from resonances, 
using Eqs.~(\ref{dminusa}) to (\ref{dva}), we find
\begin{align}
\Phi^>_-\sim&\left(\frac1r-\frac1{a_s}\right)\begin{bmatrix}1\\0\end{bmatrix}+\lambda\left(-\frac{2 u_p}{3a_s r^2}+1+\frac r{3 a_s}\right)\begin{bmatrix}0\\1\end{bmatrix},\label{pmshort}\\
\Phi^>_+\sim&\left(\frac1r-\frac2{a_s}\right)\begin{bmatrix}1\\ 0\end{bmatrix}
+\left(\frac1{\lambda r^2}+\frac\lambda 2-\frac r{\lambda u_p}\right)\begin{bmatrix}0\\ 1\end{bmatrix}.\label{ppshort}
\end{align}
At the first set of resonances given by Eq.~(\ref{cri}), 
\begin{eqnarray}
\Phi^>_-=\Phi^>_+\sim\begin{bmatrix}  1/r-2/r_0 \\ 1/\lambda r^2+\lambda/2-2\lambda r/3r_0\end{bmatrix};\label{sr1}
\end{eqnarray}
at the second set
\begin{eqnarray}
\Phi^>_-=\Phi^>_+\sim\begin{bmatrix}  1/r+3/2\lambda^2r_0^3 \\ 1/\lambda r^2+\lambda/2+r/2\lambda r_0^3\end{bmatrix},\label{sr2}
\end{eqnarray}
Note that Eqs.~(\ref{pmshort}) to (\ref{sr2}) satisfy Eq.~(\ref{sch2}) with $E=0$ to the leading order of $\lambda$.

\section{Bound States}\label{bound}
The bound state solutions to Eq.~(\ref{sch2}) with energy $E$ are Eq.~(\ref{in}) for $r<r_0$ and 
\begin{align}
\Phi^>(r)=
 C'\begin{bmatrix}h_0^{(1)}( k_1'r)\\h_1^{(1)}( k_1'r) \end{bmatrix}+ D'\begin{bmatrix}h_0^{(2)}( k_2'r)\\h_1^{(2)}( k_2'r) \end{bmatrix}
\end{align}
for $r>r_0$. Here $ k_1'\equiv\lambda+i\kappa$, $k_2'\equiv\lambda-i\kappa$, and $\kappa\equiv\sqrt{-mE}$. The requirement that the wave-function and its first derivative are continuous at $r=r_0$ gives rise to the equation determining the energy $E$:
\begin{align}
0=&(\tilde\lambda^2+\tilde\kappa^2+\tilde\kappa)(\tilde q^2-\tilde\kappa^2)[j_0(\tilde q_1)j_1(\tilde q_2)-j_0(\tilde q_2)j_1(\tilde q_1)]\nonumber\\
&+2\tilde\lambda\tilde\kappa \tilde q[j_0(\tilde q_2)j_1(\tilde q_1)+j_0(\tilde q_1)j_1(\tilde q_2)]\nonumber\\
&-2\tilde\kappa \tilde q[(\tilde\kappa+1)^2+\tilde\lambda^2]j_0(\tilde q_1)j_0(\tilde q_2)\nonumber\\
&-2\tilde\kappa\tilde q(\tilde\lambda^2+\tilde\kappa^2)j_1(\tilde q_1)j_1(\tilde q_2),\label{long}
\end{align}
with $\tilde\kappa\equiv\kappa r_0$ and $\tilde q\equiv qr_0$. Assuming $D'=1$, we have
\begin{widetext}
\begin{align}
C'\equiv e^{2i\theta}=-\frac{2\tilde qh_1^{(2)}(\tilde k_2)j_0(\tilde q_1)j_0(\tilde q_2)-h_0^{(2)}(\tilde k_2)[j_0(\tilde q_1)j_1(\tilde q_2)(\tilde q-i\tilde\kappa)+j_0(\tilde q_2)j_1(\tilde q_1)(\tilde q+i\tilde\kappa)]}
{2\tilde qh_1^{(1)}(\tilde k_1)j_0(\tilde q_1)j_0(\tilde q_2)-h_0^{(1)}(\tilde k_1)[j_0(\tilde q_1)j_1(\tilde q_2)(\tilde q+i\tilde\kappa)+j_0(\tilde q_2)j_1(\tilde q_1)(\tilde q-i\tilde\kappa)]}.\label{widec}
\end{align}
\end{widetext}
In the weak attraction limit $\eta\to 0$, we find 
\begin{equation}
\tilde\kappa=\eta^2\left[1-\left(\frac{\sin\tilde\lambda}{\tilde\lambda}\right)^2\right];\label{fbound}
\end{equation}
there is always a bound state no matter how weak the attraction is, in contrast to the case in the absence of the spin-orbit coupling. Furthermore, in the limit $\tilde\lambda\ll1$, Eq.~(\ref{fbound}) gives the bound state energy $E=-a_s^2\lambda^4/m$, agreeing with a previous pseudopotential calculation \cite{vijay2}. Equation (\ref{fbound}) fits the numerical results given in Ref.~\cite{cui}. The energies of bound states other than in the weak attraction limit have also been studied in Ref.~\cite{cui}.

According to Eq.~(\ref{long}), a bound state emerges at threshold $E=0$ when Eq.~(\ref{bcon})
is satisfied. In the limit $\tilde\lambda\to0$, Eq.~(\ref{bcon}) reduces to $j_0(\eta)=0$ or $j_1(\eta)=0$.
In the absence of the spin-orbit coupling, the former condition is where a bound state emerges in the $p$-wave channel, while the latter is where 
the $s$-wave scattering length $a_s$ is zero. The introduction of the spin-orbit coupling has the nonperturbative effect of shifting the onset of a series of bound states from where $1/a_s=0$ to where $a_s=0$. For small $\tilde\lambda$,  
the critical values of the depth of square well potential $V_c$ where a bound state forms at the zero energy threshold are given by
\begin{align}
\eta_c\equiv\sqrt{mV_cr_0^2}=\begin{cases} n\pi-\tilde\lambda^2/n\pi+\mathcal {O}(\tilde\lambda^4),\\
x_\ell+\tilde\lambda^2/x_\ell+\mathcal {O}(\tilde\lambda^4)\label{cri},
\end{cases}
\end{align}
with $n=1,2,3,\dots$ and nonzero $x_\ell$ ($\ell=1,2,3,\dots$) being the solutions of $j_1(x_\ell)=0$.
The change of $V_c$ due to small $\lambda$ agrees with the numerical results obtained in Ref.~\cite{cui}.

To obtain the asymptotic behavior of the wave-function of the most shallow bound state for $r\gtrsim r_0$ when $V_0$ is close to a critical value $V_c$, let us first look at the limit
$\eta\to\eta_c=0$ where there is only one bound state. From Eq.~(\ref{fbound}), in the limit $\tilde \lambda\to0$, for the bound state, we have $\tilde \kappa=\eta^2\tilde\lambda^2/3$. Substituting this into Eq.~(\ref{widec}) and 
expanding in the order $0<\tilde\kappa\ll \eta \ll \tilde\lambda \ll1$, we obtain
\begin{equation}
\theta=2\arctan\left({\frac{\tilde\kappa}{\tilde\lambda}}\right)\big[1+\mathcal {O}(\tilde\lambda^2)\big]\simeq2\frac{\tilde\kappa}{\tilde\lambda},
\end{equation}
and thus for $r\gtrsim r_0$ the wave-function of the bound state asymptotically is
\begin{align}
\Phi^>(r)\sim\begin{bmatrix}1/r-1/a_s\\ -2\lambda r_0^2/15 r^2+\lambda\end{bmatrix},
\end{align}
which coincides with the $\eta\to0$ limit of Eq.~(\ref{pmshort}).

When $V_0$ approaches one of the first set of critical values of Eq.~(\ref{cri}), 
\begin{equation}
\theta\simeq-\frac{1}{2}\tilde\lambda,
\end{equation}
and 
\begin{eqnarray}
\Phi^>(r)\sim\begin{bmatrix}  1/r-2/r_0 \\ 1/\lambda r^2+\lambda/2\end{bmatrix},
\end{eqnarray}
which agrees with Eq.~(\ref{sr1}) to the corresponding order of $r$. For $V_0$ close to the second set of critical values of Eq.~(\ref{cri}),
\begin{equation}
\theta\simeq \frac{2}{3}\tilde\lambda^3
\end{equation}
with
the asymptotic wave-function
\begin{eqnarray}
\Phi^>(r)\sim\begin{bmatrix}  1/r+3/2\lambda^2r_0^3 \\ 1/\lambda r^2+\lambda/2\end{bmatrix},
\end{eqnarray}
which is the same as Eq.~(\ref{sr2}) to the corresponding order of $r$.

\section{discussion}
Equations (\ref{pmshort}) and (\ref{ppshort}) indicate that away from resonances the modification of the asymptotic forms of the wave-functions at short range due to the spin-orbit coupling is perturbative. To see this point clearly, we multiply Eq.~(\ref{ppshort}) by $\lambda$ and have
\begin{align}
\Phi_-^>\sim&\left(\frac1r-\frac1{a_s}\right)\begin{bmatrix}1\\0\end{bmatrix}+\lambda\left(-\frac{2u_p}{3a_s r^2}+1+\frac r{3 a_s}\right)\begin{bmatrix}0\\1\end{bmatrix},\label{pmshort2}\\
\Phi_+^>\sim& \left(\frac1{ r^2}-\frac r{u_p}\right)\begin{bmatrix}0\\ 1\end{bmatrix}+\lambda\left(\frac1r-\frac2{a_s}\right)\begin{bmatrix}1\\ 0\end{bmatrix}
;\label{ppshort2}
\end{align}
apart from certain factors, to lowest order of $\lambda$, $\Phi^>_-$ have the same asymptotic form as the $s$-wave-function $\psi_s$ and $\Phi^>_+$ as the $p$-wave-function $\psi_p$.
The perturbative modification can be understood by inspecting the wave-functions inside the range of the potential. For $r<r_0$, in the limit $k\to0$ and expanded to the first order of $\tilde\lambda$,
\begin{align}
\Phi_-^<(r)=\Phi_1^<(r)+\left[1+4\left(\frac1\eta-\cot\eta\right)\tilde\lambda\right]\Phi_2^<(r),\label{is2}
\end{align}
and
\begin{align}
&\Phi_+^<(r)\nonumber\\
&=\Phi_1^<(r)-\left[1+\frac43\left(\frac{3-\eta^2}\eta+\frac{\eta^2}{\eta-\tan\eta}\right)\tilde\lambda\right]\Phi_2^<(r).\label{is1}
\end{align}
Thus to zero order of $\tilde\lambda$, $\Phi_+^<(r)$ and $\Phi_-^<(r)$ are the same as the $s$- and $p$-waves inside the square well potential in the absence of the spin-orbit coupling respectively. Since the coefficients of the terms linear in $\lambda$ are well-behaved away from the resonances, nonzero $\tilde\lambda$ adds corrections perturbatively. Given that $\Phi^>(r)$ and $\Phi^<(r)$ need to connect at $r=r_0$, the change of the asymptotic form of $\Phi^>(r)$ is bound to be perturbative in $\tilde\lambda$ consequently. 

The above analysis is essentailly the same as the previous argument based on the comparison of energy scales given in Ref.~\cite{zhenhua}: Within the potential range, the spin-orbit coupling strength is much smaller than $1/r_0$ and $\sqrt{mV_0}$ and therefore generally has perturbative effects on the wave-functions in this regime, and so does on the wave-functions in the regime $r\gtrsim r_0$. According to this general consideration, we expect the same conclusion applies to the cases with nonzero pair net momentum $\mathbf K$ and $\mathbf J\neq0$, and to the ones with generic forms of spin-orbit coupling, given additionally $K\ll 1/r_0, \sqrt{mV_0}$. 
If one takes the zero range limit $r_0\to0$ while with $a_s$ fixed and the scattering volume $u_p\to0$, to lowest order, Eq.~(\ref{ppshort2}) becomes the noninteracting form and Eq.~(\ref{pmshort2}) reduces to $\sim(1/r-1/a_s)[1,0]^T+\lambda[0,1]^T$, the same as found for zero range interactions in Ref.~\cite{peng}.

Nevertheless, Eqs.~(\ref{is2}) and (\ref{is1}) signal that the perturbation expansions in $\tilde\lambda$ break down when $a_s\to0$ ($\eta-\tan\eta=0$) and $1/u_p\to0$ ($\tan\eta=0$ for $\eta\neq0$), which are the conditions for the emergence of a new bound state in the limit $\tilde\lambda\to0$. At the first set of resonances, the asymptotic forms (\ref{sr1}) suggests that $\tilde\lambda$ can still be treated perturbatively. At the second set of resonances, extra powers of $\lambda$ in the denomenators of (\ref{sr2}) become singular as $\lambda\to0$ there. This coincides with the perturbative and nonperturbative shifts of the onset of bound states compared to the cases without the spin-orbit coupling mentioned in Sec.~\ref{bound}.

Our calculations suggest that for generic spin-orbit coupling, no matter for bosons or fermions, the corrections to the asymptotic form of the wave-functions at short range should be perturbative except for at fine tuned points where resonances occur. Consequently the relations derived based on the asymptotic form \cite{tan,braaten,shizhong} should stand valid correspondingly.

\begin{acknowledgments}
We thank Xiaoling Cui, Peng Zhang and Qingrui Wang for helpful discussions. This work is supported in part by Tsinghua University Initiative Scientific Research Program, and NSFC under Grant Numbers 11104157 and 11204152.

\end{acknowledgments}

\appendix

\section{Elements of the $M$ matrix}
\begin{widetext}
We list the elements of the $M$ matrix here:
\begin{align}
M_{11}=&\frac{e^{-2 i(\eta-\tilde\lambda)} \eta (\eta + \tilde\lambda)^2-e^{2 i (\eta + \tilde\lambda)} \eta (\eta - \tilde\lambda)^2 + 
   4 i \tilde\lambda [ 
      \eta^2 (i + \tilde\lambda)-\tilde\lambda^3]}{2 \eta \tilde\lambda^2 \{\eta[\cos(2 \eta) - \cos(2 \tilde\lambda)] + (\eta^2 - \tilde\lambda^2) \sin(
     2 \eta)\}},\\
M_{12}=&-i\frac{2\tilde\lambda^2(2\eta^2-\tilde\lambda^2)\cos(2\eta)-2\eta^2\tilde\lambda^2\cos(2\tilde\lambda)+\eta[(2\eta^2-1)\tilde\lambda^2-2\tilde\lambda^4-\eta^2]\sin(2\eta)+2\eta^2\tilde\lambda\sin(2\tilde\lambda)}{\eta \tilde\lambda^2 \{\eta[\cos(2 \eta) - \cos(2 \tilde\lambda)] + (\eta^2 - \tilde\lambda^2) \sin(
     2 \eta)\}},\\
M_{21}=&M_{12},\\
M_{22}=&-M_{11}^*.
\end{align}

\end{widetext}


\end{document}